\documentclass[11pt,twoside]{article}
\usepackage{./asp2014}
\newcommand{\zav}[1]{\left(#1\right)}
\newcommand{\hzav}[1]{\left[#1\right]}
\newcommand{\szav}[1]{\left\{#1\right\}}

\resetcounters

\begin{document}

\title{The Concept of Few-Parameter Modelling of Eclipsing Binary
and Exoplanet Transit Light Curves}
\author{Zden\v{e}k Mikul\'a\v{s}ek$^{1}$, Miloslav Zejda$^1$,
Theodor Pribulla$^2$, Martin Va\v{n}ko$^2$, Shen-Bang Qian$^3$,
and Li-Ying Zhu$^3$}
\affil{$^1$Department of Theoretical Physics and Astrophysics,
Masaryk University, Kotl\'a\v{r}sk\'a 2, 611 37 Brno, Czech Republic; \email{mikulas@physics.muni.cz}}
\affil{$^2$Astronomical Institute of the Slovak Academy of Sciences,
059 60 Tatransk\'a Lomnica, Slovak Republic}
\affil{$^3$Yunan Observatories, Kunming, China}
\paperauthor{Sample~Author1}{Author1Email@email.edu}{ORCID_Or_Blank}{Author1 Institution}{Author1 Department}{City}{State/Province}{Postal Code}{Country}
\paperauthor{Sample~Author2}{Author2Email@email.edu}{ORCID_Or_Blank}{Author2 Institution}{Author2 Department}{City}{State/Province}{Postal Code}{Country}
\paperauthor{Sample~Author3}{Author3Email@email.edu}{ORCID_Or_Blank}{Author3 Institution}{Author3 Department}{City}{State/Province}{Postal Code}{Country}

\begin{abstract}
We present a new few-parameter phenomenological model of light curves of eclipsing binaries and stars with transiting planets  that is able to fit the observed light curves with the accuracy better than 1\% of their amplitudes. The model can be used namely for appropriate descriptions of light curve shapes, classification, mid-eclipse time determination, and fine period analyses.
\end{abstract}

\section{Introduction}
Modern physical models of eclipsing binaries and transiting exoplanets
(eclipsing systems = ES) \citep[e.g.][]{wilson,wilson14,had,prsa,prsa1}
are able to simulate their light curves with impressive fidelity.
However, a solution of the reverse problem, to derive all important parameters of the model from observational data, is seldom unique. If we have observational data of
moderate or poor quality, we are often forced to diminish the number of
free parameters and thus use simplified (and frequently physically inconsistent) models. A decision which parameters to fix or leave floating is usually nontrivial.

Nevertheless, for several common practical tasks, such as:
1)~determination of the mid-eclipse times, 2)~light curve fitting, 3)~description and classification of ES light curve shapes for the purpose of the current and future surveys like ASAS or GAIA, and 4)~fine period analysis, we do not need to know the detailed physics of the system. A good approximation of the shape of observed light curves is usually sufficient in such cases.

We present here a few-parameter general phenomenological model of eclipsing system light curves in the form of the special, analytic, periodic functions that are able to fit an overwhelming majority of the curves with an accuracy better than 1\%. For the sake of simplicity, we will limit our considerations only to systems with approximately circular orbits that represent more than 80\% of observed eclipsing systems.

\section{The Model of Monochromatic Light Curves of Eclipsing
Systems}\label{monoch}

The model function of a monochromatic light curve (expressed in magnitudes) of  eclipsing systems $F(\varphi,\lambda)$, can be assumed as the sum of three particular functions:
\begin{equation}
F(\varphi,\lambda)=F_{\mathrm e}(\varphi,\lambda)+F_{\mathrm p}(\varphi,\lambda)+F_{\mathrm c}(\varphi,\lambda),
\end{equation}
where $F_{\mathrm e}$ describes the mutual eclipses of the components,
$F_{\mathrm p}$  models the proximity effects,
while $F_{\mathrm c}$ approximates the O'Connell effect
(irrespectively of its physical cause).

\begin{figure}[ht]
\centering \resizebox{0.5\textwidth}{!}{\includegraphics{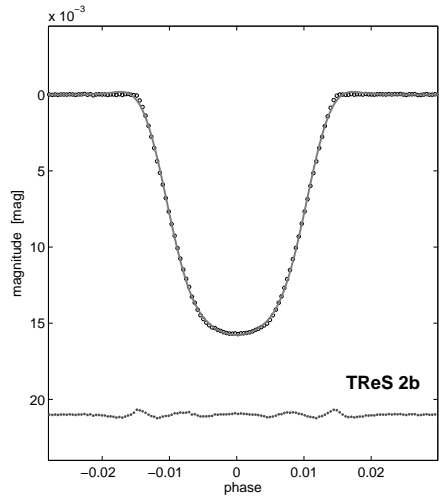}}
\begin{center}
\caption{\small The {\it Kepler} light curve of an exoplanet transit in TrES-2b \citep{treska} is represented by points while the model fit is shown by a line.
The differences between the observed magnitudes and the model,
magnified by a factor of 10, are shown in the lower part of the
figure.}\label{Fig1}
\end{center}
\end{figure}

The normally occurring features of all eclipsing system light curves are two more or less symmetrical depressions caused by mutual eclipses of gravitationally bound stellar or planetary components. The model function of eclipses $F\!_{\mathrm e}$ can be approximated by the sum of two symmetrical analytic functions of the phase $\varphi$. We assume that the primary eclipse is centered at the phase $\varphi=0$, while the secondary eclipse (if one is present) is centered at phase $\varphi=0.5$,
\begin{eqnarray}\label{monokl}
&\displaystyle F\!_{\mathrm e}(\varphi)=\sum_{k=1}^{n_{\mathrm e}}
A_k\hzav{1\!+\!C_{k}\zav{\frac{\varphi_k}{D_k}}^2+\!K
\zav{\frac{\varphi_k}{D_k}}^4 }\szav{1\!-\!\szav{1\!-\!\exp\hzav{1\!-\!
\cosh\zav{\frac{\varphi_k}{D_k}}}}^{\mathit{\Gamma}_{\!k}}};\\
& \varphi_k=\hzav{\varphi-(k-1)/2} -
\mathrm{round}\hzav{\varphi-(k-1)/2},\nonumber
\end{eqnarray}
where the summation is over the number of eclipses, $n_{\mathrm e}$:
$n_{\mathrm e}=2$ or $n_{\mathrm e}=1$ (the common situation for exoplanet transits), $\varphi_k$ are the auxiliary phases defined in the interval $\langle -0.5,0.5 \rangle$, $A_k$ are the central depths of eclipses in mag, $D_k$ is the parameter expressing half-widths of eclipses, $C_k, K$ are the correcting parameters, $\mathit {\Gamma}_{\!k}$  parameterize the kurtosis of individual eclipses.

The fit of a transit of the exoplanet TrES2b ($n_{\mathrm e}=1$) observed by {\it Kepler} (see Fig.\,\ref{Fig1}) \citep{treska} utilizes five parameters:
$A_1,C_1,K,D_1,\mathit {\Gamma}_1$. However, the full description of the model monochromatic light curves with two eclipses ($n_{\mathrm e}=2$) would generally require nine free parameters: $A_{1,2},C_{1,2},D_{1,2},\mathit {\Gamma}_{1,2}$, and $K$. Nevertheless, the most common situations would permit a reduction to 5 parameters, $A_{1,2},C,D,\mathit {\Gamma}$, because it is usually acceptable to assume that
$C_1\simeq C_2$, $K\simeq 0$, $D_1\simeq D_2$, $\mathit {\Gamma}_1\simeq \mathit {\Gamma}_2$. A detailed discussion of the properties of the eclipse model
function $F\!_{\mathrm e}$ and its possible simplifications are given in \citet{basic}.

Light variations of eclipsing binaries caused by eclipses are usually modified by
proximity and O'Connell effects that may be modelled by a sum of cosine and sine harmonic polynomials:
\begin{equation}\label{prox}
\displaystyle F\!_{\rm p}(\varphi)= \sum_{k=n_{\mathrm e}+1}^{n_{\mathrm p}+
n_{\mathrm e}} A_k\cos\hzav{2\,\pi\,(k\!-n_{\mathrm e})\,\varphi};\quad
F_{\rm c}(\varphi)= \sum_{k=n_{\mathrm p}+n_{\mathrm e}+1}^{n_{\mathrm c}+
n_{\mathrm p}+n_{\mathrm e}} A_k\,\sin(2\,\pi\,\varphi),
\end{equation}
where $k=n_\mathrm{e}+1,n_\mathrm{e}+2,\ldots,
(n_{\mathrm e}+n_{\mathrm p}+n_{\mathrm c})$, $n_{\mathrm p}$
is the number of terms in $F\!_{\rm p}(\varphi_1)$: $n_{\mathrm p}=0,2,3,\ldots$, $n_{\mathrm c}=0$ if the O'Connell asymmetry is not present\footnote{If $p>q$ then $\sum_{k=p}^q\,h_k=0.$}, else $n_{\mathrm c}=1$
\citep[for more information see][]{basic}.

\begin{figure}
\centering \resizebox{0.6\hsize}{!}{\includegraphics{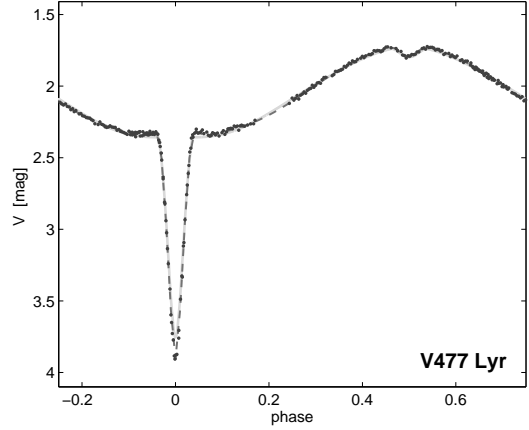}}
\caption{\small{$V$ light curve of an unusual eclipsing binary V477~Lyr
consisting of a very hot nucleus of a planetary nebula and solar type star.
Only eight free parameters is needed for the fit of the light curve
strongly affected by proximity effects. $V$ data taken from \citet{pol}.}} \label{477lyr}
\end{figure}

The uncommon $V$ light curve of eclipsing binary V477\,Lyr (see Fig.\,\ref{477lyr}) is determined by 8 parameters: $A_{1,2,3,4,5},D,C,\mathit{\Gamma}$, $n_{\mathrm p}=3,\,n_{\mathrm c}=0$.

\section{Multicolour Light Curves of Exoplanet Transits and
Eclipsing Binaries}

The parameters of the above defined model functions, especially
the amplitudes $A_k(\lambda)$, and parameters $C_k(\lambda)$,
$D(\lambda),$ and  $\mathit{\Gamma}(\lambda)$ are generally
wavelength dependent.
\begin{figure}[ht]
\centering \resizebox{0.6\hsize}{!}{\includegraphics{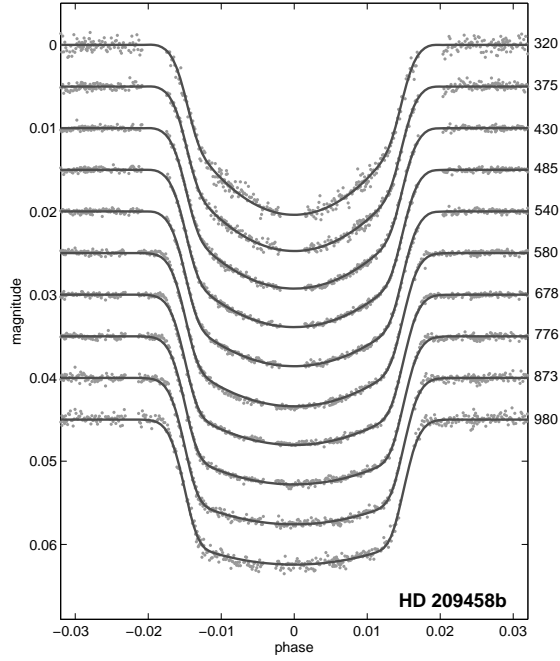}}
\begin{center}
\caption{\small Light curves of the 10-colour photometry of the
exoplanet transit of HD\,209458b. Effective wavelengths of individual
colours are in nm. The data were taken from \citet{knut}} \label{Fig0}
\end{center}
\end{figure}
It follows both from the theory and experience that most of the wavelength dependencies can be well approximated by the low order polynomials of the quantity
$\mathit {\Lambda}$; $\mathit{\Lambda}=\lambda_0/\lambda_{\mathrm {eff}}-1$, where
$\lambda_{\mathrm {eff}}$ is the effective wavelength of corresponding
passband, $\lambda_0$ is an arbitrarily selected central wavelength of
the data set. Then:
\begin{equation}\label{multi}
\displaystyle A_{k}=\sum_{j=1}^{g_{ {ak}}}a_{kj}\,\mathit{\Lambda}^{j-1},\ \
C_l=\sum_{j=1}^{g_{\mathrm {c}l}}c_{lj}\,\mathit{\Lambda}^{j-1},\ \
D=\sum_{j=1}^{g_d}d_{\!j}\,\mathit{\Lambda}^{j-1},\ \
\mathit{\Gamma}=\sum_{j=1}^{g_{ \gamma}}\gamma_k\,\mathit{\Lambda}^{j-1},
\end{equation}
where $g_{ak}, g_{cl},g_{d}, g_{\gamma}$, $l=1,2$ or $l=1$,
are the numbers of degrees of freedom of the corresponding
parameters of the model. The standard set of the monochromatic
light curve model parameters of eclipsing binaries (see Sec.\,\ref{monoch}):
$\{A_k,C_l,D,\mathit{\Gamma}\}$ can be considered as the special
case of the multicolour decomposition Eq.\,\ref{multi}:
$\{a_{k1},c_{l1},d_1,\mathit{\gamma}_1\}$.

\begin{figure}[ht]
\centering \resizebox{0.6\hsize}{!}{\includegraphics{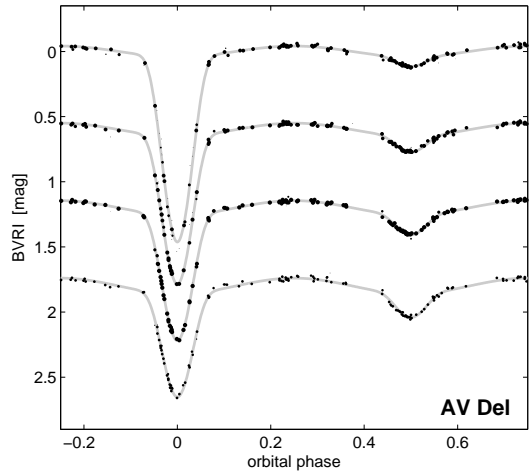}}
\begin{center}
\caption{\small \textit{BVRI} light curves of a `cool Algol' AV~Del,
consisting of a main sequence F type primary and the K subgiant
filling its Roche lobe, can be approximated by our model function with 9 free parameters. The data were taken
from \citet{mader}. } \label{Fig3}
\end{center}
\end{figure}

All needed parameters and their uncertainties can be determined
using the standard non-linear weighted least square technique
\citep[see in][]{basic}.

\section{Conclusions}

The outlined phenomenological modeling of eclipse-binary
and planetary-transit light curves has an advantage of high
efficiency and simplicity and thus can be used for a wide range
of applications.

\acknowledgements The study is supported by the grant of Ministry
of Education of the Czech Republic LH14300.


\begin{thebibliography}  

\bibitem[Hadrava(2004)]{had} Hadrava P.\ 2004, Publ. Astron.
Inst. Czechosl. Acad. Sci., 92, 1

\bibitem[Knutson et al.(2007)]{knut} Knutson H. A., Charbonneau D.,
Noyes, R. W. et al. 2007, \apj, 655, 564

\bibitem[Mader et al.(2005)]{mader} Mader J. A., Torres, G.,
Marschall L. A., \& Rizvi A. 2005, \aj, 130, 234

\bibitem[Mikul\'a\v{s}ek \& Zejda, 2013]{mikzej} Mikul\'a\v{s}ek Z. \&
Zejda M., in \'Uvod do studia prom\v{e}nn\'{y}ch hv\v{e}zd,
ISBN 978-80-210-6241-2,  Masaryk University, Brno 2013

\bibitem[Mikul\'a\v{s}ek et al.(2008)]{mik901} Mikul\'a\v{s}ek Z.,
Krti\v{c}ka J., Henry G. W., Zverko J., \v{Z}i\v{z}\v{n}ovsk\'y et al.
2008, \aap, 485, 585

\bibitem[Mikul\'a\v{s}ek(2015)]{basic} Mikul\'a\v{s}ek Z. 2015,
\aap, submitted

\bibitem[Pollacco \& Bell(1994)]{pol} Pollaco D. L. \& Bell S. A. 1994,
\mnras, 267, 452

\bibitem[Pr{\v s}a \& Zwitter(2005)]{prsa} Pr{\v s}a A., \& Zwitter T. 2005, \apj, 628, 426

\bibitem[Prsa et al.(2011)]{prsa1} Prsa, A., Matijevic, G.,
Latkovic, O. et al. 2011, Astrophysics Source Code Library, 6002

\bibitem[Raetz et al.(2014)]{treska} Raetz S., Maciejewski G., Ginski C., Mugrauer M. et al. 2014, \mnras, 444, 1351

\bibitem[Wilson \& Devinney(1971)]{wilson} Wilson R.~E., \&
Devinney, E.~J.\ 1971, \apj, 166, 605

\bibitem[Wilson \& Van Hamme(2014)]{wilson14} Wilson R.~E., \& Van Hamme W.\ 2014, \apj, 780, 151

\end{thebibliography}
\end{document}